\documentclass[11pt,a4paper]{article}
\usepackage{times}
\usepackage{url}
\usepackage{latexsym}
\usepackage{epsfig}
\usepackage{amssymb}
\usepackage{mathtools}
\usepackage{tabularx}
\usepackage{amssymb}
\usepackage{amsmath}
\usepackage{bm}
\usepackage{breakcites}

\newcolumntype{E}{>{\small}c}
\newcolumntype{C}{>{\small}X}
\newcolumntype{S}{>{\hsize=.3\hsize}C}

\newcommand{\U}{\mathbf{U}}
\newcommand{\V}{\mathbf{V}}

\title{Sub-Story Detection in Twitter with Hierarchical Dirichlet Processes}

\author{
  { \bf P. K. Srijith$^{1}$, Mark Hepple$^{1}$, Kalina Bontcheva$^{1}$, Daniel Preotiuc-Pietro$^{2}$} \\
  $^{1}$ Department of Computer Science, The University of Sheffield, UK \\
  $^{2}$ Computer \& Information Science, University of Pennsylvania, USA \\
  {\tt \{pk.srijith,m.r.hepple,k.bontcheva\}@sheffield.ac.uk}\\ {\tt danielpr@sas.upenn.edu}
}

\date{}

\begin{document}
\maketitle
\begin{abstract}
Social media has now become the de facto information source on real world events. The 
challenge, however, due to the high volume and velocity nature of social media streams,
is in how to follow all posts pertaining to a given event over time -- a task referred to 
as story detection. 
Moreover, there are often several different stories pertaining to a given event, 
which we refer to as \emph{sub-stories} and the corresponding task of their automatic
detection -- as \emph{sub-story detection}.  
This paper proposes hierarchical Dirichlet processes (HDP), a probabilistic  topic model, as an effective method for automatic sub-story detection. HDP can learn sub-topics associated with sub-stories which enables it to handle subtle variations in sub-stories. It is compared with state-of-the-art story detection approaches based on locality sensitive hashing and spectral clustering. 
We demonstrate the superior performance of HDP for sub-story detection  on real world Twitter data sets using various evaluation measures. The ability of HDP to learn sub-topics helps it to recall the sub-stories with high precision.
Another contribution of this paper is in demonstrating that 
the conversational structures within the Twitter stream can be used to
improve sub-story detection performance significantly.

{\bf keywords :} sub-story detection, hierarchical Dirichlet process, spectral clustering, locality sensitive hashing
\end{abstract}

\section{Introduction}
\label{sect:intro}

Online social networks play a major role in generating and disseminating information. They provide a platform for people to voice their opinion and viewpoints. Moreover, social media provide main stream media, governments, companies, and citizens the opportunity to obtain real time information about events happening around the world. The 
challenge, however, due to the high volume and velocity nature of social media streams,
is in how to follow all posts pertaining to a given event over time -- a task referred to 
as \emph{story detection} \cite{sasa10}. 

Story detection is a specific form of Topic Detection and
Tracking (TDT) \cite{Allan2002}, which is concerned with discovering posts pertaining to real-world stories as they unfold over
time. This paper, in particular, is concerned with tracking event-related posts within the
micro-blogging social network, Twitter. 

When considering posts related to a given real-world event (e.g. the Ferguson unrest), 
researchers (e.g. \cite{zubiaga2016plos,procter2013reading}) have found that there are often different \emph{sub-stories} pertaining that event
(e.g. `Ferguson police shot Brown', `Ferguson police confiscated all video
evidence'). In this paper we address the question how such substories can be identified automatically. This is particularly 
beneficial for mainstream media, governments, law enforcing agencies, companies, and other organisations who are 
increasingly faced with the challenge of tracking and responding to emerging sub-stories, rumours, and misinformation. 

Automatic sub-story detection is a harder task than story detection as
sub-stories on the same event have a lot of vocabulary in common. For instance, all the posts related to the Ferguson unrest share words such as `Ferguson', `police', `Brown'. Sub-stories tend to overlap considerably in time  and are often associated with low tweet rates. Also, they are not necessarily true (e.g. different rumours emerged simultaneously during the England riots \cite{procter2013reading}) and thus may not have a corresponding sub-story in the real world.  Standard story detection (also known as event detection) approaches fail to take these into account and do not perform well on sub-story detection task.  


This paper investigates a sub-story detection approach based on hierarchical Dirichlet processes (HDP). 
This hierarchical topic model can detect
sub-topics associated with a main topic and is thus well suited to the sub-story
detection task. For instance, in the Ferguson unrest example, the
approach can detect `Ferguson police' as the main topic, while `shot Brown'
and `video evidence' as different sub-topics. The effectiveness of the proposed approach is established by comparing it with state-of-the-art story
detection approaches based on locality sensitive hashing (LSH)~\cite{sasa10}
and spectral clustering (SC)~\cite{Preotiuc2013} (see Section~\ref{sect:methods} for details).

Sub-story detection in Twitter streams is particularly challenging, since 
reply tweets often do not share sufficient vocabulary with the
tweets they reply to. For instance, in one of the tweet data set used for experiments
here \cite{zubiaga2016plos}, there is a sub-story claiming that Fox News is not
covering the Ferguson protests. In particular, one of the 
earlier tweets has text `Currently the \#FoxNews website
has zero, repeat, ZERO coverage of the \#Ferguson protests'. 
A subsequent reply tweet, however, states
``Too busy bitching about POTUS for sure'', i.e. does not mention neither 
Ferguson, nor Fox News explicitly. Since many reply tweets exhibit similar 
lack of linguistic overlap with the originating tweet, this makes it difficult to 
cluster them as pertaining to the same sub-story cluster. To address this 
problem, our approach takes into account the conversational structure, in addition 
to linguistic features, in order to improve sub-story detection performance.

Sub-story detection performance is evaluated on
real world Twitter data sets.  Twitter is ideal for this task because
it provides a large source of publicly available posts on major
events, e.g.\ the 2014 Ferguson unrest and the 2011 London riots. 
In particular, we experiment with five different Twitter data sets. 
Three of these arise from rumour detection research 
\cite{zubiaga2016plos,procter2013reading} and come annotated with 
events (2011 London riots, the 2014 Ferguson unrest, 
and the 2014 Ottawa shooting) and tweets grouped by sub-story within each of 
the events. The fourth public data set was created for first story 
detection (FSD) with Locality Sensitive Hashing (LSH), which is one of our baselines~\cite{sasa10}. 
The fifth public data set (FAcup~\cite{luca13}) contains tweets about various events occuring 
during a football match.  
The properties of the data sets are discussed further in 
Section~\ref{sect:data set} and comparative evaluation results for all methods and data sets are
detailed in Section~\ref{sect:experiments}. 

We propose to use a mutual information based evaluation measure, adjusted mutual information (AMI)~\cite{vinh09}, in addition to the standard precision-recall
metrics which avoids agreements by chance by considering the number of clusters produced. The experimental results establish the superior performance of HDP  for sub-story detection task.
The approach could detect sub-story specific topics which helps journalists and government agencies to monitor the evolution of new topics associated with a story.  
The runtime performance of HDP is comparable to established story detection approaches and can be used to perform real-time detection of sub-stories.

The main contributions of the paper are summarized as follows:
\begin{itemize}
 \item Introduces the sub-story detection task and proposes a hierarchical Dirichlet process based approach to solve the problem of sub-story detection.
 \item Provides rigorous experimental comparison of the proposed sub-story detection approach with state-of-the-art story detection approaches, establishing the effectiveness of HDP for sub-story detection.
 \item Proposes a mutual information based metric for evaluating the performance of sub-story detection approaches.
 \item Demonstrates the usefulness of  conversational structure in improving  sub-story detection performance. 
\end{itemize}

\section{Related Work}
\label{sect:related}

A number of techniques have been employed for detecting and tracking
stories in social media streams~\cite{Allan2002}. Story detection is typically done by
extending traditional clustering algorithms to a streaming data
setting~\cite{aggarwal14}. A comprehensive survey of the literature on
story detection techniques in Twitter data is given in
\cite{Farzindar2013}.

Clustering approaches were traditionally used to group data points
with similar features \cite{jain99}. Many of the classic clustering
techniques such as K-means have been extended to a streaming data
setting~\cite{zhong05, aggarwal10} and can be used to perform story
detection~\cite{aggarwal14}. These algorithms aim to discover
underlying groups within data by inferring a general representation to
characterize the data in terms of a few key topics or stories. 

Story
detection in Twitter for a particular topic such as `earthquake' is
studied in \cite{sakaki10}. Becket {\it et al.}~[2011] use an online clustering algorithm to detect
stories and distinguish real vs.\ non-real stories using a
classification method. Twevent~\cite{li12} is a story detection approach which
clusters bursty segments in Twitter data. A fast and efficient approach based on
locality sensitive hashing (LSH) is first used in \cite{sasa10} to detect the
emergence of new stories (first story detection) in Twitter. Locality sensitive hashing reduced the computational complexity associated with nearest neighbor 
search and detected clusters of documents in constant space and time. Later, they extended this approach to counter 
lexical variations in documents by using paraphrases~\cite{sasa12}. An alternative approach to detect new events by storing the contents
of already seen documents in a single hash table is proposed in \cite{wurzer15b}. Further, LSH based techniques are also developed
to handle topic streams emerging in Twitter~\cite{wurzer15a}. Here, topics are also hashed into a bucket in addition to the tweets.

Topic models are also
 used to detect stories in Twitter, for instance  latent Dirichlet allocation (LDA)~\cite{diao12} to detect trending topics.
 A non-parametric topic model based on Dirichlet process   is
used in \cite{wang13} to detect newsworthy stories in Twitter, where
topics are shared among tweets from consecutive time periods. Topicsketch~\cite{xie13} uses a novel sketch based topic model to 
detect bursty topics from millions of tweets.

A spectral clustering based approach is developed in 
\cite{preotiuc16} to address the task of story detection. The approach uses a mutual information based metric to represent the similarity
matrix in spectral clustering. Real world events happen at different scales of time and space. Multi-scale event detection~\cite{dong15} 
aims to detect stories evolving at different pace and spanning different geographic locations by using the properties of wavelet transform. 
Supervised machine learning techniques such as support vector machines and logistic regression are used to detect events corresponding
to specific topics such as those related to traffic~\cite{andrea15}, lifestyle and wellness~\cite{akbari16} and uprisings~\cite{ben15}. These event detection
techniques will not be able to distinguish different sub-stories associated with a main story due to content overlap.
 
Whilst story  detection has received considerable attention, less
attention has been paid to sub-story detection task. Aiello { \it et al.}~[2013] discuss tasks similar to sub-story detection like finding important events happening over time in a main event such as a football match. They used standard story detection approaches to find events on their tasks. There exists approaches~\cite{nichols12, zubiaga12}  which rely on tweet rates in an interval to detect major moments in a game. Chakrabarti and Punera \cite{punera11} use a modified hidden
Markov model combining tweet rate and text features to
summarize events in a game. Shen { \it et al.}~[2013] use a
time-content mixture model which effectively combines burstiness and
cohesiveness to detect key moments in a story. Chierichetti { \it et al.}~[2014] use non-textual features based on tweet rate and 
communication pattern among users to detect points in time where an important event happens within a story.  An approach based on graph-of-words to represent sequence of tweets was used in 
\cite{melad15} to detect important events happening within a football match. 
These approaches will not be effective for detecting sub-stories which overlaps considerably in time and have low tweet rates.  
\nocite{chier14}

There exists a hitherto unaddressed problem -- finding sub-stories related to a particular real world event. These sub-stories may or may not correspond to real-world events (e.g. false rumours about the London riots \cite{procter2013reading} do not), they tend to overlap in time (i.e. tweets on more than one sub-story circulate simultaneously), and share significant common vocabulary \cite{zubiaga2016plos}. As demonstrated in the rest of this paper, state-of-the-art approaches for story detection do not perform well on this type of task. 

\section{Research Objective}

The following are our main research objectives:

\begin{enumerate}
 \item Introduce the task of sub-story detection in Twitter. Sub-story detection differs from story detection and we discuss the properties specific to sub-story detection which makes it a harder task than story detection.    
 \item Propose hierarchical Dirichlet processes as an effective approach for sub-story detection. Unlike story detection approaches, HDP can learn sub-topics associated with sub-stories which makes it  particularly useful for modeling sub-story detection task.
 \item Verify experimentally the effectiveness of HDP for sub-story detection task. We compare HDP with standard story detection approaches based on locality sensitive hashing and spectral clustering on real world Twitter data sets to establish the fitness of HDP for sub-story detection. 
 \item Show the usefulness of conversational structure in Twitter for improving the performance of sub-story detection task. By considering conversational structure, reply tweets which does not share a topical similarity with the source tweets gets clustered  along with the source tweet.
 \item Introduces adjusted mutual information score as an effective metric to measure clustering performance in sub-story detection. Standard metrics based on  precision typically favor clustering approaches which produces large number of small sized clusters. Such clustering approaches are not useful in practice and we propose to use AMI as an effective alternative metric which can take care of such problems.
\end{enumerate}



\section{Problem Definition : Sub-Story Detection}
\label{sect:problem}


This paper addresses the problem of detecting \emph{sub-stories} as they emerge in social media streams. Automatic sub-story detection methods need to separate tweets related to different sub-stories into different clusters, even though they pertain to the same real-world event.   

Sub-story detection
differs from a {story\/} detection in that sub-stories share
some common vocabulary and the tweet rates for the sub-stories are
comparatively low. Table~\ref{tab:fer} shows 8
major sub-stories related to the Ferguson unrest from one of our five data sets. 
All these sub-stories are related to the shooting of M. Brown by the Ferguson police and thus share words such as `M. Brown', `Ferguson', `police' etc.   Standard story detection approaches fail to produce good results in this setting where vocabulary is shared across the sub-stories because they look at tweet similarity or overlaping words to cluster tweets.  

\begin{figure}[!htbp]
\vspace*{-2mm}
\begin{center}
\includegraphics[scale=0.45]{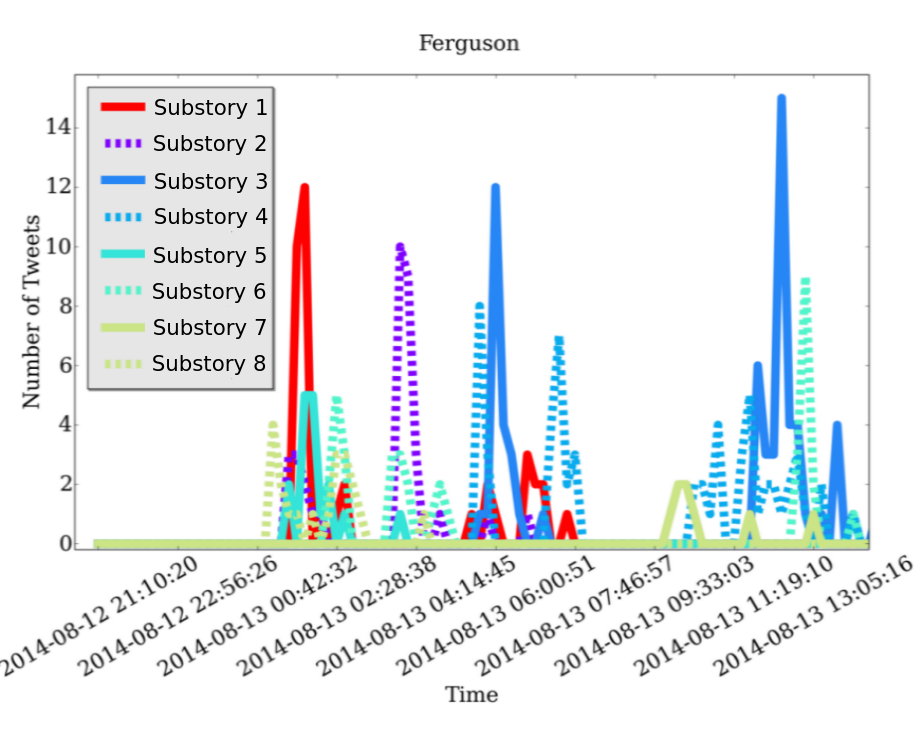}
\vspace*{-8mm}
\caption{Temporal profile of sub-stories in the Ferguson data set}
\label{fig.time}
\end{center}
\vspace*{-3mm}
\end{figure}

\begin{table}[!htbp]
\caption{Description of major stories in the Ferguson data set.}
\label{tab:fer}
\vspace{1mm}
\begin{tabularx}{\textwidth}{|S|C|S|S|}
\hline
Sub-story id  & Description & \# tweets & \# source \\
\hline
1 &  M. Brown was involved in a robbery before being shot & 2312 & 89 \\
2 &  Initial contact between police officer and M. Brown was not related to the robbery & 553 & 28 \\
3 &  Ferguson police to release name of police officer who shot M. Brown today (August 15) & 536 & 26 \\
4 &  Ferguson police are leading a smear campaign or character assassination of M. Brown & 623 & 18 \\
5 &  Ferguson police once beat up a man and charged him for bleeding on their uniforms & 369 & 24 \\
6 &  Ferguson police are lying about the circumstances leading up to M. Brown's death & 426 & 23 \\
7 &  M. Brown was stopped by police for walking in the middle of the street & 236 & 10 \\
8 &  Fox News is not covering the Ferguson protests & 102 & 2 \\
\hline
\end{tabularx}
\vspace{-4mm}
\end{table}

Sub-stories can have  considerably larger lifespan, overlap in time, and are set within a broader over-arching story, that contains many thematically related sub-stories.  The themes discussed in these sub-stories are referred to as \emph{sub-topics}.  For example, consider the temporal profile of sub-stories from the Ferguson data shown in
Figure~\ref{fig.time}.~\footnote{These sub-stories are not the same as those listed in
Table~\ref{tab:fer}.} We can observe that these sub-stories overlap in time and have relatively low tweet rate. This is mainly due to the fact that there are multiple conversations  within a sub-story, each evolving at a different point in time. A sub-story which becomes active at some point in time can become dormant temporarily and then re-activate again at a later time (for instance, consider the temporal behavior of sub-stories 1 and 3 in Figure~\ref{fig.time}).

\section{Data Set Description}
\label{sect:data set}

The core of our experiments are carried out on three sub-story annotated data sets (Ferguson unrest, Ottawa shooting and London riots). 
The first two are very recent and have been collected and human annotated as part of a rumour analysis project~\cite{zubiaga2016plos}, while the London riots one arose from an earlier qualitative social science analysis of related tweets~\cite{procter2013reading}. All three data sets consist of tweets, grouped together into human annotated
sub-stories related to the particular real-world event. Other tweets pertaining to the same event count as background data. 

We also consider two other publicly available data sets, FSD and FAcup, which
have been created for story detection. Even though these data sets are not strictly suitable to the sub-story detection task, we use them for comparative purposes and  have vocabulary overlap across some sub-stories.  

Next we describe these data sets in more detail. 

\subsection{Ferguson}

This data set consists of tweets collected between August and September
2014, all related to the unrest that took place in Ferguson, USA~\cite{zubiaga2016plos}. 
This data set is not only sub-story annotated, but also includes ``reply-to''
information, which connects together subsets of tweets into
conversational graphs. We refer to {\it reply tweets\/} as those
that reply to a tweet present in the data
set, and {\it source tweets\/} as all those which do not have such a ``parent''. 
In other words, the data set is regarded as a collection of conversational threads,
each of which has a single source tweet at its root.  

As detailed in~\cite{zubiaga2016plos}, journalists categorised manually source tweets
as belonging to one of 45 different sub-stories.  A reply tweet is assumed automatically to 
belong to the sub-story to which
its source tweet has been assigned (if any). 

After discounting sub-stories with
fewer than 10 tweets, the final data set consists of 6,598  labeled tweets
spread across 35 sub-stories and 18,650
tweets as background, i.e. not belonging to any of those sub-stories. 
Considering source tweets alone, there are 284
labeled source tweets and 899 background source
tweets. Table~\ref{tab:fer} lists major sub-stories in the
data~\cite{zubiaga2016plos}, illustrating how the sub-stories are very
similar and have the Ferguson unrest as a common topic.

\subsection{Ottawa}

The Ottawa data set consists of tweets related to shootings at the
parliament building in Ottawa during October 2014~\cite{zubiaga2016plos}. 
Similar to the Ferguson data,
it also has a conversational structure including source and reply tweets.  
The data set consists of 6,414 tweets spread
across 39 sub-stories and 5,975  tweets as background. 

Considering source tweets alone, there are 462
labeled tweets and 439 background tweets. Some major sub-stories associated with the Ottawa shooting event are `Soldier shot dead is Nathan Cirillo', `Soldier shot at War Memorial has died', `Suspected shooter is dead', etc. All these sub-stories have a common theme of shooting and death which makes them an ideal candidate for the sub-story detection task. With respect to temporal patterns, the evolution of sub-stories is very similar to the patterns observed in the Ferguson data.   

\subsection{London Riots}

The London riots data set consists of 2.5 million tweets related to
the riots that took place in London during August, 2011~\cite{procter2013reading}. It includes
10,000 tweets that are labeled as belonging to 7 different
sub-stories, all with a common background topic -- the London riots. 
Table~\ref{tab:lon} provides a summary of number of tweets in each  sub-story in this data set. 
Unlike Ferguson and Ottawa, the conversational structure was not made available by the researchers.

\begin{table}[t]
\caption{Description of sub-stories for London riots data set.}
\label{tab:lon}
\vspace{10pt}
\centering{
\begin{tabular}{|c|c|c|}
\hline
Sub-story id  & Description & \# tweets\\
\hline
1 & Army deployed in Bank & 192 \\
2 & Rioters attack children's hospital  & 1666 \\
3 & London Eye set on fire & 657 \\
4 & Rioters cook food in McDonalds & 218 \\
5 & Miss Selfridge set on fire & 5581 \\
6 & Police beat girl & 902 \\
7 & Rioters attack London zoo & 937 \\
\hline
\end{tabular}
}
\end{table}

\subsection{First Story Detection}

This is a publicly available story detection data set~\cite{sasa12} with approximately 
2,400 tweets labeled as belonging to 27
stories, from the period June to September 2011. This is augmented
with background tweets from the same period, to create a corpus of
approximately 80,000 tweets. Originally this data set was created by
Petrovi\'{c} { \it et al.}~[2012] for evaluation of
their first story detection (FSD) system. This FSD data set
can be seen to represent the standard { story\/} detection task, in contrast to the { sub-story\/}
task represented by the former three data sets. It should be noted, however,
that there is { some\/} overlap of stories here as well, e.g.  four of
the stories are related to the London riots in 2011 and another four are about death of some celebrities. 
These commonalities make this data also applicable to sub-story detection, as well as enabling us to 
benchmark our methods on the related story detection problem.

\subsection{FAcup}
This is a publicly available  data set~\cite{luca13} with approximately 7,000 tweets belonging to 13 different sub-stories associated with a football match story. These tweets represents sub-stories such as goals, fouls etc. in the 2012 Football Association (FA) final match between Chelsea and Liverpool. This data set is augmented with approximately 20,000 tweets related to the same football match as background. Due to the shared common story (football game), sub-story tweets in this data set share a common vocabulary and is useful for evaluating the proposed sub-story detection approach. However, they differ from sub-story data sets such as Ferguson and Ottawa in that the sub-stories in this data set are temporally separated.     

\section{Methods}
\label{sect:methods}

The main sub-story detection method investigated in this paper uses
hierarchical topic modeling. In particular, we experiment with hierarchical 
Dirichlet processes (HDP), a
non-parametric Bayesian model, which can effectively model the
sub-story detection task. HDP is also compared to two story detection 
state-of-the-art approaches: spectral clustering and locality
sensitive hashing.

\subsection{Hierarchical Dirichlet Process}
\label{sect:hdp}

Latent Dirichlet allocation (LDA)~\cite{blei03}
and hierarchical Dirichlet processes (HDP)~\cite{teh06} have shown
promising results in topic modeling due to their probabilistic
interpretations. They model a document (i.e. tweet in our case) as a mixture of topics,
where each topic has a distinct distribution over the words. These
generative models can infer the latent topics associated with the tweets.  

In this paper we propose to use HDP for sub-story detection, since
it can model the hierarchical structure underlying the topic
distribution. As argued above, in sub-story detection we need to find sub-topics
associated with the main story (e.g. the Ferguson unrest), and HDP is
developed specifically to handle such kinds of tasks. 
HDP achieves this by extending the Dirichlet process mixture model (DPMM)~\cite{pml} to a hierarchical setting. 

In more detail, the DPMM considers a tweet as 
consisting of words generated by a mixture of topics. The mixture distribution is
modeled using a non-parametric prior based on a Dirichlet process (DP)~\cite{nils10}. 
A DP is parameterized by a base distribution $H$ and a concentration parameter 
$\alpha$ and is denoted as $DP(\alpha, H)$.  The base measure specifies the 
a-priori distribution over some parameter space $\bm \theta$ which is used 
to generate observed data. 

In our case, $\bm \theta$ represents the parameters of a multinomial distribution 
over the words $\bm w$ in a tweet.  A draw from $DP(\alpha, H)$ is a discrete 
probability measure $G$ providing a distribution over $\bm \theta$. It can be 
represented as $G(\bm \theta) = \sum_{i=1}^{\infty} \pi_i \delta_{\theta_i}(\bm \theta)$, 
where $\delta_{\theta_i}$ is the Kronecker delta function which gives a value 
of $1$ when the parameter takes value $\theta_i$, $\theta_i$ is a draw from 
$H$ and $\pi_i$ is the probability mass associated with $\theta_i$. The 
sequence of values $\pi_i$ is obtained from $\alpha$ using a stick breaking process~\cite{sethu94}

\begin{equation}
\pi_i = \bar{\pi}_i \prod_{l=1}^{i-1}(1-\bar{\pi}_l)   \qquad \qquad   \bar{\pi}_i \sim Beta(1, \alpha).
\end{equation}

The process ensures that $\bm \pi$ represents a probability distribution 
\textit{i.e.} $\sum_{i=1}^\infty \pi_i = 1$ and is often represented as $\bm{\pi} \sim GEM(\alpha)$. 
The concentration parameter $\alpha$ determines the probability mass associated 
with a topic $\pi_i$ as a parameter in Beta distribution.
A draw from  $G$ results in $\theta_i$ with probability $\pi_i$, with $\theta_i$ 
representing the parameters of a multinomial distribution associated with a topic $i$. 
Thus each topic $i$ occurs in a tweet with probability $\pi_i$. 
Modeling tweets independently as a DPMM  does not allow topics to be shared across 
tweets, which is needed in our task. 

Hierarchical Dirichlet processes are developed to handle grouped 
data and share topics across the groups~\cite{teh06}. We use them to 
model the tweet as consisting of a set of topics  and to share topics across multiple tweets. 
HDP achieves this by drawing tweet specific probability distribution $G_d$  for a tweet $d$ 
from $DP(\gamma, G_0)$, where $\gamma$  represents the concentration parameter and 
$G_0$ is the base distribution shared by all the tweets. The common base distribution 
$G_0$ is indeed a draw from $DP(\alpha, H)$. The common base distribution has the form 
$G_0(\bm \theta) = \sum_{i=1}^{\infty} \pi_{0i} \delta_{\theta_i}(\bm \theta)$ and 
the tweet specific distribution has the form 
$G_d(\bm \theta) = \sum_{i=1}^{\infty} \pi_{di} \delta_{\theta_i}(\bm \theta)$. Here, both the common base distribution and tweet specific distributions share the 
parameters $\theta_i$ (or the topics) with tweet specific mixture distribution $\bm \pi_d$ over 
the topics. Thus, the tweets modeled
using HDP share the topics but with different probabilities. The mixing proportions 
$\bm \pi_d$ is generated as follows~\cite{teh06}
\begin{equation}
\bm{\pi_d} \sim DP(\gamma, \bm{\pi_0})    \qquad \qquad    \bm{\pi_0} \sim GEM(\alpha) .
\end{equation}

\begin{figure}[!htbp]
\vspace*{2mm}
\begin{center}
\includegraphics[scale=0.6]{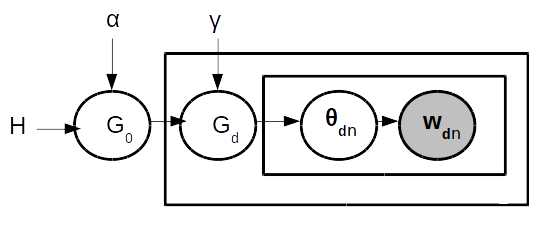}
\vspace*{-2mm}
\caption{Graphical model for HDP}
\label{hdp-graphical-model}
\end{center}
\vspace*{-6mm}
\end{figure}

Figure~\ref{hdp-graphical-model} shows the graphical model representation of the HDP model. 
A word ${w}_{dn}$ in a tweet $d$ comes from a topic with  parameter $\theta_{dn}$ 
drawn from the Dirichlet Process $G_d$ associated with the tweet. The topics are shared 
across the tweets due to the hierarchical modeling of DPMM.

Since in sub-story detection tweets relate to
the same real world event, HDP can model this effectively, 
coupled with the fact that individual tweets address different 
sub-topics (corresponding to the sub-stories). 
These sub-topics are characterized by words  and each word is associated 
with a probability indicating the importance of the word in 
representing the sub-topic.  The identified sub-topics are  used to
cluster tweets based on the words common in the tweet and 
the sub-topics.  For each tweet, we detect common words and 
calculate a similarity score to a sub-topic by summing the 
probability associated with these  words in representing 
the sub-topic. The tweet is assigned to the sub-topic with the maximum 
similarity score. We use sub-topics for clustering the tweets 
as they can better discriminate the tweets associated with sub-stories.

\subsection{Spectral Clustering}
\label{sect:sc}

Clustering techniques have been used widely to detect stories in social
data streams \cite{aggarwal14}. Here, we discuss one based on spectral
clustering using point-wise mutual information~\cite{Preotiuc2013}. 
The spectral clustering (SC) algorithm~\cite{shi00}  has been shown to achieve the
state-of-the-art performance for tasks such as community detection in
graphs~\cite{smyth05}.  This method treats clustering as
a graph partitioning problem. It projects the objects  into a
lower dimensional  space by performing singular value decomposition of a similarity graph constructed over the objects. It then discovers clusters of objects which are maximally
separated in this space using standard clustering techniques, such as k-means. A good spectral clustering relies on a good similarity graph which best reflects the connections between objects.


We apply spectral clustering to detect sub-stories in a stream
of tweets~\cite{Preotiuc2013}. The approach represents a similarity graph by constructing a matrix which captures the similarity over words
that appear in the data. It uses normalized point-wise mutual information
(NPMI) \cite{bouma09} to capture the word similarity. NPMI measures the
probability of co-occurrence of words in the same tweet. The idea is that if two
words appear consistently in the same tweet, then they are indicative of the same story. For example, the
co-occurrence of words such as `Ferguson' and `police' over a period of time indicate there is a
story related to Ferguson police. 

The NPMI measure between words pairs $x$ and $y$ is calculated as
\begin{equation}
 NPMI(x,y) = -\log p(x,y)  \log \frac{p(x,y)}{p(x)p(y)}
\end{equation}
where $p(x)$ denotes the probability of occurrence of a word $x$ in a tweet, 
and $p(x,y)$ provides the probability of co-occurrence of words $x$ and $y$ in a tweet. 
We consider two words as co-occuring if they appear in the same tweet, which 
gives us a straightforward measure of co-occurence frequencies. The NPMI 
measure takes values between $-1$ and $1$ with positive values 
indicating a higher chance of co-occurrence and negative value indicating 
a lower chance of co-occurrence. 


The spectral clustering algorithm proceeds by filtering out less frequent words and constructing a similarity graph over words using the NPMI measure. It ignores all  NPMI values less than a threshold and keeps the largest connected component from the resulting graph. Singular value decomposition is performed over a graph Laplacian constructed from this similarity graph to obtain a representation of words in a lower dimensional space. A k-means algorithm then finds clusters of words in this reduced space.

The word clusters discovered by the spectral clustering  algorithm 
represents a coherent topic. The words are associated with a score, which
provides a measure of importance of the word in representing the
topic. For each tweet, a similarity score is computed with respect to
each topic, using the co-occurence score of the words 
in the tweet. The tweets
are then clustered by assigning them to the topic
with the highest similarity score. Thus, tweets in the same cluster form a
topically coherent cluster.

\subsection{Locality Sensitive Hashing}

The second state-of-the-art approach is locality sensitive hashing
(LSH)~\cite{raja11}, which was proposed originally for first story 
detection in Twitter \cite{sasa10}. LSH finds nearest neighbour tweets in constant time and keeps only
a constant number of tweets in memory. 

The LSH approach uses the nearest neighbour algorithm to
find the tweet closest to the incoming tweet. The computational
overhead of finding the nearest neighbor is overcome using  locality sensitive hashing. LSH maps incoming tweets to buckets using a
hashing function which maps similar tweets to the same bucket. 
The method then finds the nearest neighbour to the incoming tweet 
by searching the bucket to which it has
been mapped. This greatly reduces the search space. The approach 
clusters tweets based on the cosine similarity of the tweets which 
are hashed into the same bucket. It assigns an incoming tweet to the 
cluster of its nearest neighbour if the cosine similarity is greater 
than a particular threshold. Otherwise, it assigns the tweet to a new cluster.  

In more detail, LSH uses a series of random hyperplanes sampled from a normal Gaussian distribution. 
These hyperplanes divide the space into subspaces and similar tweets fall 
into the same subspace. We consider $k$ such hyperplanes.  The number of 
hyperplanes $k$ can be considered as a number of bits per key in this hashing 
scheme. Let $\bm u_i, i=1\ldots k$ represent the hyperplanes and $\bm x$ be the 
{\it tf-idf} representation of the tweet. The hash value is considered to be a 
binary vector with $k$ bits. We set the bit $i$ to be 1 if $\bm x.\bm u_i > 0$ 
and 0 otherwise. The tweets falling in the same subspace have the same hash 
value in the hash table and is stored in a bucket of size $b$.  The higher $k$ is,  
the fewer collisions there will be in the buckets, but more time will be needed to 
compute the hash values. However, increasing k also decreases the probability of 
collision with the nearest neighbor, and hence multiple hash tables ($h$) 
are required to increase the chances of finding the nearest neighbor. Thus, a tweet 
is compared with the tweets belonging to the same bucket in multiple hash 
tables in order to find its nearest neighbor using cosine similarity. The nearest neighbor tweets with cosine similarity below a user specified threshold forms a cluster. This cluster represents tweets with some topical similarity, which helps one to detect stories evolving in Twitter.  

\section{Experimental Results}
\label{sect:experiments}


This section reports on the comparative evaluation of HDP and all state-of-the-art baselines on the sub-story detection data sets. We follow a cluster-based approach, as it accounts for the varying popularity of sub-stories and the related user endorsement aspect. This also provides a fair comparison with LSH. Alongside this, we also consider an event extraction setting where major sub-stories detected by HDP and SC are described in terms of detected topics\footnote{Please note that LSH does not assign topics to sub-stories, due to the nature of the algorithm.}. Results are reported using the standard metrics of precision, recall, F-score and adjusted mutual information~\cite{vinh09}. The latter is included as it has certain advantages over the others with respect to cluster evaluation.  

In particular, the experiments compare Hierarchical Dirichlet processes (HDP), spectral clustering (SC) and locality sensitive
hashing (LSH) on the data sets introduced in Section~\ref{sect:data set} (Ferguson, Ottawa, London riots, FSD, and FA Cup). 

The text of each tweet is pre-processed to remove unusual characters, tokens, and stop
words, followed by stemming.  In particular, the filtered
tokens are: user-mentions (tokens starting with @), hashtags (words starting with \#) and URLs. 
The rationale behind hashtag removal is that hashtags often tend to refer to the shared real world event (e.g. \#Ferguson, \#Londonriots) and are thus shared across sub-stories.

\subsection{Method Comparison using Precision, Recall and F-score}

Our first comparative evaluation experiment uses the standard information retrieval metrics of precision and recall~\cite{manning08}. Detected tweet clusters are evaluated against the gold standard tweets in the respective sub-stories. Since the approaches are unsupervised, the number of automatically discovered clusters does not always align to the sub-stories in the gold standard.  Therefore, for each sub-story, we find the automatically produced cluster with the maximum overlap, in terms of number of tweets from that sub-story. It should be noted that multiple sub-stories may get aligned to a single cluster. In this case, precision measures how many of the retrieved tweets belong to the actual sub-story, while recall measures whether the system could retrieve all known tweets associated with the aligned sub-story. Performance is reported using micro-averaged precision and recall, due to the varying sizes of each sub-story (in terms of number of contained tweets).

More formally, let $N$ be the total number of known sub-stories in a given data set. $TP_i$, $FP_i$, and $FN_i$ are the true positives, false positives, and false negatives associated with a sub-story $i$. Then, micro-averaged precision and recall are calculated as:
\begin{eqnarray}
P_{micro} = \frac{\sum_{i=1}^N TP_i}{\sum_{i=1}^N TP_i + FP_i} \\
R_{micro} = \frac{\sum_{i=1}^N TP_i}{\sum_{i=1}^N TP_i + FN_i}
\end{eqnarray}

We also report micro-averaged F-score, which is the harmonic mean of micro-averaged precision and recall. Approaches with a high F-score are preferred.
\begin{equation}
F_{micro} =  \frac{2 \cdot P_{micro} \cdot R_{micro}}{P_{micro} + R_{micro}}
\end{equation}


The free HDP parameters (e.g. the concentration parameters) are learnt from the data using Gibbs sampling. We put an upper bound on the number of topics produced by HDP ($k$) and this allows for a fair comparison with SC.  The effective number of topics could be less and is determined by the concentration parameter learnt from the data.The spectral clustering approach depends primarily on the parameter $k$, which determines the number of clusters in the data set. The approach is run by filtering out words with  an NPMI threshold of 0.1 and with word frequency threshold of 10. We perform experiments  with different values of $k$ for HDP and SC. Lastly, the LSH approach depends on the parameters $k$ (number of bits), $h$ (number of hash tables) , and $b$ (bucket size). The experiments are conducted with different values of these parameters. We present only the results obtained with the best two parameter settings (in terms of F-score) for each of the approaches.

As can be seen in Table~\ref{tab:expt2}, HDP is the best performing method on the Ferguson and Ottawa sub-story data sets. In particular, HDP's F-score is significantly better than the SC and LSH F-scores. With respect to precision, LSH performs best, while SC has the best recall. The low recall of LSH, however, is due to the fact that it generates a large number of very small tweet clusters, which is also the reason for its high precision. On the other hand, SC is not able to differentiate sufficiently between similar sub-stories and groups them together in a small number of very large clusters. While this increases recall, it leads to the observed low precision. In contrast, HDP can detect subtle differences in sub-stories, thanks to the sub-topics, which are then used to cluster the tweets. This leads to improved precision for HDP over SC and an ultimately higher F-score. 

\begin{table*}[!htbp]
\caption{Results of HDP, SC, and LSH  on Ferguson and Ottawa data sets for different parameter settings. Best results appear in . }
\label{tab:expt2}
\vspace{10pt}
\centering{
\begin{tabular}{|c|c|c|c|c|c|c|}
\hline
 & \multicolumn{3}{c|}{Ferguson} &  \multicolumn{3}{c|}{Ottawa} \\
 \hline
 Method & $P_{micro}$ & $R_{micro}$ & $F_{micro}$ & $P_{micro}$ & $R_{micro}$ & $F_{micro}$  \\
\hline
HDP (k200) & 0.0536 & 0.0889 & 0.0668 & 0.1799  & 0.1431 & \bf{0.1594} \\
HDP (k300) & 0.1366 & 0.1057 & \bf{0.1191} &  0.2182 & 0.1249 & 0.1588 \\
 \hline
SC (k400)  & 0.0131  & \bf{0.1622} & 0.0242 & 0.0519 & \bf{0.1821} & 0.0807 \\
SC (k2000) & 0.0422  & 0.0861 & 0.0566 & 0.0873 & 0.1244 & 0.1025 \\
\hline
LSH (k12h56b10)  & \bf{0.3441}  & 0.0301 & 0.0554   & \bf{0.4797} & 0.0314 & 0.0589 \\
LSH (k13h71b10)  & 0.3430  & 0.0407 & 0.0728 & 0.3768 & 0.0285 & 0.0529 \\

\hline
\end{tabular}
}
\end{table*}

Next, Table~\ref{tab:exptlon} reports the experimental results on the much larger London riots data set.
The methods here are executed by partitioning the data set into 50 sub-partitions
with approximately 50,000 tweets in each. The table shows
the number of clusters per partition for the HDP and SC approaches. As above, HDP has the best recall and F-score, while LSH has the highest precision.

\begin{table*}[!htbp]
\caption{Results of HDP, SC, and LSH on London riots  for different parameter settings. Best results are indicated in bold letters. }
\label{tab:exptlon}
\vspace{10pt}
\centering{
\begin{tabular}{|c|c|c|c|c|c|c|c|}
\hline
&  \multicolumn{3}{c|}{London Riots} \\
 \hline
 Method & $P_{micro}$ & $R_{micro}$ & $F_{micro}$ \\
\hline
 HDP (k50) & 0.4188  & \bf{0.2759} &  \bf{0.3326}\\
 HDP (k100) &  0.4194 & 0.2013 &  0.2720\\
\hline
 SC (k50) &  0.1833  & 0.2666 & 0.2172 \\
 SC (k100) & 0.4522   & 0.2539  & 0.3252\\
\hline
 LSH (k12h56b10) & \bf{0.5948}   & 0.2258 & 0.3273  \\
 LSH (k13h71b10) & 0.4976  & 0.2323 & 0.3167  \\

\hline
\end{tabular}
}
\end{table*}

Nevertheless, it should be noted that precision, recall and F-score are still very low on the Ferguson and Ottawa data sets, which is due to the presence of conversational threads within the sub-story clusters. As discussed in Section~\ref{sect:intro}, some of the tweets within the conversational tweets tend to discuss completely unrelated topics. For instance, even if the source tweet mentions the sub-story explicitly, often reply tweets within the thread do not have significant word overlap with the source. Consequently, these reply tweets are not deemed topically similar to the source tweet and are assigned to a completely different cluster, which negatively impacts performance. 

\subsubsection{Conversational Structure Experiments}
\label{sect:conversations}

The effect of reply tweets in lowering the performance of the system is verified by conducting clustering experiments on the Ferguson and Ottawa data sets, using source tweets alone.  As can be seen in Table~\ref{tab:sourcetweets}, algorithm performance improves significantly, compared to the results reported in Table~\ref{tab:expt2}. In some cases, the  improvement in performance is by an order of magnitude. Again we observe that HDP outperforms both LSH and SC, with similar precision and recall patterns as those observed in the full data sets.   

\begin{table}[!htbp] 
\caption{Results of HDP, SC, and LSH   on Ferguson and Ottawa data sets (considering source tweets alone) for different parameter settings. Best results are indicated in bold letters. .} 
\label{tab:sourcetweets}
\vspace{10pt}
\centering{
\begin{tabular}{|c|c|c|c|c|c|c|}
\hline
 & \multicolumn{3}{c|}{Ferguson} &  \multicolumn{3}{c|}{Ottawa} \\
 \hline
 & $P_{micro}$ & $R_{micro}$ & $F_{micro}$ & $P_{micro}$ & $R_{micro}$ & $F_{micro}$  \\
 \hline
HDP (k200) & 0.2578 & 0.3042 & 0.2790 & 0.4615  & 0.4172 & 0.4382 \\
HDP (k300) & 0.3482 & 0.3688 & \bf{0.3582} &  0.4212 & \bf{0.5248} & \bf{0.4673} \\
\hline 
SC (k400)  & 0.0508 & \bf{0.3691} & 0.0893 & 0.1574 & 0.3055 & 0.2077 \\
SC (k2000)  & 0.1614 & 0.2617 & 0.1996 & 0.1373  & 0.3055 & 0.1894 \\
\hline
LSH (k12 h56 b10) & \bf{0.6083} & 0.2402 & 0.3444  &  0.6417 & 0.2603 & 0.3703 \\
LSH (k13 h71 b10)  & 0.5591 & 0.2508 & 0.3462  & \bf{0.6975} & 0.2451 & 0.3627 \\
\hline
\end{tabular}
}
\end{table}

The next experiment considers the sub-story assignment of entire conversational threads. The first step is to cluster only the source tweets, while reply tweets are assigned automatically to the cluster of their corresponding source tweet. This is a realistic setting on these data sets, and on unseen Twitter data in general, since the source-reply structure is readily available.

Table~\ref{tab:expt3} shows that considering conversational threads achieves an order of magnitude improvement in recall and F-score, as compared to those in Table~\ref{tab:expt2}. Again, we observe that LSH has better precision, while HDP has better recall, and ultimately HDP has the best F-score. 

This experiment also considered an additional baseline, which clusters tweets using only the conversational structure. By design, this approach has a precision of 1. The aim here is to investigate whether the sub-story detection approaches can get better recall, than this readily available baseline. This is indeed the case, as shown in Table~\ref{tab:expt3}. In particular, the simple baseline has a recall of $0.2545$ and $0.1696$ for Ferguson and Ottawa respectively, which is lower than the recall of the three other methods.


\begin{table*}[!htbp]
\caption{Results of HDP, SC, and LSH   on Ferguson and Ottawa data sets (considering conversational structure) for different parameter settings. Best results are indicated in bold letters. }
\label{tab:expt3}
\vspace{10pt}
\centering{
\begin{tabular}{|c|c|c|c|c|c|c|}
\hline
 & \multicolumn{3}{c|}{Ferguson} &  \multicolumn{3}{c|}{Ottawa} \\
 \hline
Method & $P_{micro}$ & $R_{micro}$ & $F_{micro}$ & $P_{micro}$ & $R_{micro}$ & $F_{micro}$  \\
\hline
HDP (k200) & 0.273 & 0.3674 & 0.3132 & 0.4749  & 0.4612 & 0.4679 \\
HDP (k300) & 0.3822 & \bf{0.4199} & \bf{0.4001} &  0.4398 & \bf{0.5691} & \bf{0.4968} \\
 \hline
SC (k400)  & 0.0722  & 0.4091 & 0.1227 & 0.1786 & 0.3581 & 0.2383\\
SC (k2000) & 0.2149  & 0.3034 & 0.2515	 & 0.1588 & 0.3143 & \bf{0.2109} \\
\hline
LSH (k12 h56 b10)  & \bf{0.5589}  & 0.3087 & 0.3977   & 0.5428 & 0.3038 & 0.3895 \\
LSH (k13 h71 b10)  & 0.5079  & 0.3106 & 0.3854 & \bf{0.7777} & 0.2877 & 0.4200 \\

\hline
\end{tabular}
}
\end{table*}

In order to investigate variation in method performance across individual sub-stories, 3 major sub-stories are selected at random in the Ferguson and Ottawa data sets. As can seen in Table~\ref{tab:expt4}, performance patterns for LSH, SC, and HDP remain unchanged, i.e. LSH  has the best precision, while HDP -- the best recall and F-score. The latter is able to find most of the tweets associated with sub-stories with a good precision. 

\begin{table*}[t]
\caption{Results of HDP, SC, and LSH  on most prominent stories in the Ferguson and Ottawa data sets.}
\label{tab:expt4}
\vspace{10pt}
\centering{
\begin{tabular}{|c|c|c|c|c|c|c|c|c|c|}
\hline
\multicolumn{10}{|c|}{Ferguson} \\
\hline
& \multicolumn{3}{c|}{Sub-story 1} &  \multicolumn{3}{c|}{Sub-story 2} &  \multicolumn{3}{c|}{Sub-story 3}\\
 \hline
Method & P & R & F & P & R & F & P & R & F \\
\hline
HDP (k300) & 0.96  & 0.26 & 0.41 &  0.45  & 0.40  & 0.42 & 0.59  & 0.74  & 0.66 \\
\hline
SC (k400) & 0.32 &  0.26 & 0.28	 &  0.16 &	0.19	& 0.17  & 0.23 & 0.29 & 0.26 \\
\hline
LSH (k12h56b10)  & 0.99	& 0.16 & 0.28 &  1  & 0.17 & 0.29  & 0.31  & 0.26 & 0.28 \\
\hline
\multicolumn{10}{|c|}{Ottawa} \\
\hline
 & \multicolumn{3}{c|}{Sub-story 1} &  \multicolumn{3}{c|}{Sub-story 2} &  \multicolumn{3}{c|}{Sub-story 3}\\
 \hline
Method & P & R & F & P & R & F & P & R & F \\
\hline
HDP (k300) & 0.99 & 0.65 & 0.78 & 0.82 & 0.66 & 0.73 & 0.46 & 0.47 & 0.46 \\
\hline
SC (k400) & 0.95 & 0.27 &	0.42  & 0.15 & 0.31 & 0.20  & 0.11  & 0.28 & 0.16  \\
\hline
LSH (k13h71b10)  & 0.99 & 0.44 & 0.61 &  1 & 0.15 & 0.26  & 0.4	 & 0.16 & 0.23 \\
\hline
\end{tabular}
}
\end{table*}

\subsubsection{Performance on FSD data}

The next experiment compares the methods on the publicly available FSD story detection data set (see Table~\ref{tab:expt1}). 
As before, LSH has very high precision but low recall. HDP and SC outperform LSH in recall, while HDP precision is better than
that obtained for SC. Thus, again HDP has the highest F-score.

Similar to the sub-story data sets, LSH produces very small clusters, which split the tweets
belonging to a particular story across multiple threads resulting in
higer precision but low recall. For instance in the case of the story
on `Death of Amy Winehouse' with 726 tweets, the corresponding LSH
cluster contains 109 tweets mostly from that story. The precision for
this story is thus $0.90$, while recall is only $0.13$.

In the FSD data, we observed that LSH produced around 1500 clusters in total, after ignoring clusters with fewer than 3 tweets.
Spectral clustering, on the other hand, tends to cluster together
tweets from related stories, resulting in few large story clusters. For example, some of
the tweets from the two stories (`Death of Amy
Winehouse' and `Betty Ford dies') are put into the same cluster.
In the case of `Death of Amy Winehouse', the
corresponding SC cluster has 821 tweets, with
$0.59$ precision and $0.67$ recall. 
This is mainly due to SC clustering words rather than messages, and thus merging sub-stories sharing common vocabulary.

HDP provides a more balanced
result with comparatively higher precision and recall. It is a more
fine grained approach which can distinguish subtle differences in
various stories, due to the hierarchical modeling of the topics with
some shared vocabulary. In the case of `Death of Amy Wine
house', the corresponding HDP cluster has 660 tweets with $0.81$
precision and $0.73$ recall.

\begin{table*}[!htbp]
\caption{Results of HDP, SC, and LSH on FSD   for different parameter settings. Best results are indicated in bold letters. }
\label{tab:expt1}
\vspace{10pt}
\centering{
\begin{tabular}{|c|c|c|c|c|c|c|c|}
\hline
 & \multicolumn{3}{c|}{FSD}  \\
 \hline
Method & $P_{micro}$ & $R_{micro}$ & $F_{micro}$  \\
\hline
HDP (k200) & 0.3181 & \bf{0.7765} & 0.4523 \\
HDP (k300) & 0.3558 & 0.7549 & \bf{0.4863} \\
\hline
SC (k200) &  0.1564 & 0.7683 & 0.2598  \\
SC (k400) & 0.1529 & 0.5406 & 0.2383 \\
\hline
LSH (k12h56b10) & \bf{0.9792} & 0.2279 & 0.3697   \\
LSH (k13h71b10) & 0.8128 & 0.2428 & 0.3739  \\

\hline
\end{tabular}
}
\end{table*}

In conclusion, this experiment has demonstrated that HDP performs very well also on story detection data sets and tasks.  

\subsubsection{Performance on FAcup Data}

The publicly available  FAcup data set is used to study the behavior of the story detection approaches. The data exhibits properties similar to Ferguson and Ottawa since all the tweets belongs to a common main event, i.e. a football match, which makes it a challenging task for the three methods. 

Table~\ref{tab:facup} compares the performance HDP, SC, and LSH  on the FAcup data.
The methods struggle to separate the tweets into the different sub-story clusters, which leads to lower precision. Broadly speaking, the results obtained on the FACup data are similar to those reported in Table~\ref{tab:expt2} on the sub-story detection data sets. Again, HDP outperforms LSH and SC, thanks to superior recall and F-score, while LSH  maintains the best precision. 

\begin{table*}[!htbp]
\caption{Results of HDP, SC, and LSH on FAcup for different parameter settings. Best results are indicated in bold letters. }
\label{tab:facup}
\vspace{10pt}
\centering{
\begin{tabular}{|c|c|c|c|c|c|c|c|}
\hline
 & \multicolumn{3}{c|}{FAcup}  \\
 \hline
Method & $P_{micro}$ & $R_{micro}$ & $F_{micro}$  \\
\hline
HDP (k100) & 0.1441 & \bf{0.1683} & 0.1552 \\
HDP (k200) & 0.3023 & 0.1438 & \bf{0.1949} \\
\hline
SC (k400) &  0.0582 & 0.0947 & 0.0721  \\
SC (k1000) & 0.1281 & 0.0875 & 0.1039 \\
\hline
LSH (k12h56b10) & 0.4975 & 0.0881 & 0.1496   \\
LSH (k13h71b10) & \bf{0.4992} & 0.0979 & 0.1636  \\
\hline
\end{tabular}
}
\end{table*}

\subsubsection{Discussion}

The experiments presented above demonstrated that LSH generally produces a large number of clusters with high precision but low recall. For instance, on the London riots data set, it produced around 45,000 clusters. In contrast, HDP and SC achieve similar performance with only 2500 and 5000 clusters, respectively. 

In general, LSH tends to create numerous very small clusters (mostly containing re-tweets), which explains its very high precision. On 
the other hand, SC tends to cluster together similar categories, which lowers precision. HDP distinguishes subtle topical differences,
resulting in more balanced precision and recall.  Another noteworthy observation is that, in general, increasing the number of clusters in HDP and SC leads to improved precision but at the cost of recall. Thus, depending on application needs, HDP and SC make it possible to trade off some recall for better precision.

With respect to the metrics used above, precision, recall and F-score do not penalize methods, such as LSH, which produce a large number of small clusters, and thus the corresponding F-score is often high due to their high precision. Such methods, however, are not useful in practice as an user has to navigate over a large number of clusters, in search of important sub-stories. Therefore, in our final experiment  we use adjusted mutual information (AMI)~\cite{vinh09}, which takes cluster size and cluster numbers into account. The improvement in clustering quality due to HDP is clearly visible also with adjusted mutual information, which corrects for agreement by chance due to a larger number 
of clusters.

\subsection{Adjusted Mutual Information Experiments}

The information theoretic, adjusted mutual information measure (AMI)~\cite{vinh09} is used to evaluate cluster quality. In prior work, information theoretic measures, such as mutual information, have been shown as being well suited to comparing the performance of clustering approaches~\cite{banarjee05, meila05}. These measures are theoretically grounded and provide a better evaluation of cluster quality. 

Mutual information (MI) between two clustering $\U = \{U_1, \ldots, U_R\}$ (true clustering of tweets) and $\V = \{V_1, \ldots, V_C \}$ (generated clustering of tweets) quantifies the information shared among them and provides the reduction in uncertainty on $\U$ upon observing $\V$. The MI score  between $\U$ and $\V$, is computed as
\begin{equation}
 MI(\U, \V) = \sum_{i=1}^R \sum_{j=1}^C p(i,j) \log \frac{p(i,j)}{p(i)p(j)}.\end{equation}
Here, $p(i)$ is the probability that tweets belong to cluster $U_i$, $p(j)$ -- the probability that tweets belong to cluster $V_j$, and  $p(i,j)$ -- the probability that tweets belong to both clusters $U_i$ and $V_j$. When clusters are identical, MI score takes a higher value upper bounded by $max\{H(\U), H(\V)\}$, where $H(\U) = -\sum_{i=1}^R p(i)\log (p(i))$ is the entropy of cluster $\U$. If the clusters are disjoint, MI score is close to zero. One can also use a normalized MI (NMI) score, which normalizes the MI score to be between zero and one.

The shortcoming of the MI and NMI scores, however, is that they do not correct for clusters that occur by chance. They do not have a constant baseline value, i.e. the average value obtained for a random clustering of the data~\cite{vinh09}. Consequently, these scores tend to be higher for results with larger number of clusters, or when the ratio of the total number of data points to number of clusters is small. In particular, they would produce a high score for an approach, which categorizes each tweet into a separate cluster. 

Therefore, in our experiments we consider adjusted mutual information (AMI) \cite{vinh09}, which is corrected for chance by subtracting the expected mutual information score from both the numerator and denominator of the normalized mutual information score. The AMI score is calculated as follows
\begin{equation}
AMI(\U, \V) = \frac{MI(\U, \V) - \mathbb{E}\{MI(\U, \V)\}}{max\{H(\U), H(\V)\}- \mathbb{E}\{MI(\U, \V)\}}.
\end{equation}

Table~\ref{tab:exptami1} and Table~\ref{tab:exptami2} provide the AMI scores obtained by HDP, SC and LSH on the different data sets. As can seen, HDP has the best performance, as measured by its AMI score. In this case, we also note that SC demonstrated improved performance and tends to be better than LSH on most data sets. As expected, the AMI score penalizes the LSH algorithm, which produces a very large number of clusters, since the expected mutual information score grows as the number of clusters increases.

\begin{table}[!htbp]
\caption{Adjusted mutual information scores for HDP, SC and LSH on the Ferguson, Ottawa, FSD and FAcup  data sets. The best AMI scores obtained for different parameter setting of the approaches are reported and the best results are shown in bold.}
\label{tab:exptami1}
\vspace{10pt}
\centering{
\begin{tabular}{|c|c|c|c|c|}
 \hline
Method  & Ferguson & Ottawa & FSD & FAcup \\
\hline
HDP (k100)   & 0.46 & 0.59 & \bf{0.70}  & 0.10 \\
HDP (k200)  & 0.46 & 0.55  & 0.67  & \bf{0.11} \\
HDP (k300)   & \bf{0.47} & \bf{0.60} & 0.65 & 0.10 \\
 \hline
SC (k200)   & 0.40 & 0.39 &  0.65  & 0.07 \\
SC (k400)   & 0.38 & 0.43 &  0.45 & 0.07 \\
SC (k2000)   & 0.39 & 0.42 &  0.31 & 0.08  \\
\hline
LSH (k12 h56 b10)   & 0.40 & 0.46   & 0.23 & 0.08 \\
LSH (k13 h71 b10)   & 0.40 & 0.47 & 0.24  & 0.09 \\
\hline
\end{tabular}
}
\end{table}

\begin{table}[!htbp]
\caption{Adjusted mutual information scores for HDP, SC and LSH on the London riots data set. The best AMI scores obtained for different parameter setting of the approaches are reported and the best results are shown in bold.}
\label{tab:exptami2}
\vspace{10pt}
\centering{
\begin{tabular}{|c|c|}
 \hline
 Method  & LondonRiots \\
\hline
 HDP (k25)  & \bf{0.32} \\
 HDP (k50)  & 0.31\\
 HDP (k100)  &0.29\\
 \hline
SC (k50)  & 0.31\\
SC (k100)  & 0.31\\
SC (k200)  & 0.28 \\
\hline
LSH (k12 h56 b10)  & 0.29\\
 LSH (k13 h71 b10) & 0.30 \\
\hline
\end{tabular}
}
\end{table}

\subsection{Topic detection}

Since HDP and SC are topic based and can describe a cluster through key terms, this is not the case for LSH. 
Therefore, in this section we examine topics within sub-stories, as identified automatically by these two methods. 

In particular, Table~\ref{tab:top} shows 5 major topics learnt by HDP and SC
from the Ferguson data. We found that HDP learns topics
corresponding to major stories in the Ferguson data set. For instance,
 Topic 1, Topic 2 and Topic 3 correspond to Story 1, Story 7 and Story 3 in the Ferguson data set, described in Table~\ref{tab:fer}. 
The first two topics detected by SC correspond to Story 5 and Story 3 of the Ferguson data.

\begin{table*}[!htbp]
\caption{Top 5 Topics identified by HDP and SC on the Ferguson data set.}
\label{tab:top}
\vspace{10pt}
\centering{
\begin{tabularx}{0.9\textwidth}{|C|C|C|C|}
\hline
\multicolumn{2}{|c|}{\scriptsize HDP} & \multicolumn{2}{c|}{ \scriptsize SC} \\
\hline
Shared & police, ferguson & Shared & None\\
\hline
Topic 1 &  suspect, robbery, brown, mike, officer, involved  & Topic 1 &  beat, charged, man, property, uniforms \\
\hline
Topic 2 & officer,  darren,  wilson,  shot,  brown,  michael & Topic 2 & brown, darren,  michael, officer, shot, wilson\\
\hline
 Topic 3 & chief,  stopped,  robbery, says,  street, walking & Topic 3 & law, live, militarized,  state,  town,  victim \\
 \hline
Topic 4&charged, beat, man, bleeding, uniforms, property, destruction & Topic 4 & before, boy, community, dogs, fergusonshooting \\
\hline
 Topic5 & store, video, stills, surveillance, robbery, brown, release, michael & Topic 5 &   mike, name, release, police \\
 \hline
\end{tabularx}
}
\end{table*}

\subsection{Runtime Efficiency}

We study the runtime of different approaches on the data sets and check their practical usability.
Table~\ref{tab:exptrt} provide runtime comparisons of HDP, LSH and SC
approaches on different data sets. The algorithms are run on a Linux
computer having 4 core Intel CPU with 3.40 GHz speed and 16 GB RAM.
In terms of run time, the performance of all the approaches are
comparable in Ferguson, Ottawa, FSD and FAcup. In the case of London riots,
LSH is found to have relatively higher runtime.

\begin{table*}[t]
\caption{Running times of  the approaches on London riots, Ferguson, Ottawa, FSD and FAcup data sets. }
\label{tab:exptrt}
\vspace{10pt}
\centering{
\begin{tabular}{|c|c|c|c|c|c|}
 \hline
Method  & LondonRiot & Ferguson & Ottawa & FSD & FAcup  \\
\hline
HDP   & 2 hours & 196 seconds & 55 seconds & 661 seconds & 188 seconds \\
 \hline
SC   & 1.5 hours & 183 seconds & 52 seconds &  550 seconds & 128 seconds\\
\hline
LSH   & 4 hours & 151 seconds & 35 seconds & 511 seconds & 152 seconds\\
\hline
\end{tabular}
}
\end{table*}

\section{Discussion and Implications}

Social networks such as Twitter provide real time information on various events happening around the world. Sub-story detection in Twitter provides news agencies and government organizations the ability to track the evolution of various stories associated with a main story. For instance, it helps journalists to detect various stories associated with U.S. presidential elections and government to track stories arising around natural disasters such as earthquakes. The proposed approach based on HDP could detect accurately most of the sub-stories associated with a main story in real time. It will be useful for news agencies and governments to more accurately track the evolution of sub-stories and take appropriate remedial measures. The sub-topics learnt by HDP from the Twitter helps humans to understand the content of sub-stories without actually inspecting them. It also avoids the need to have a separate algorithm to summarize the contents of the sub-stories. 

We take into account  the conversational structure in Twitter which allows our model to more accurately track the evolution of sub-stories. By observing the rate of growth of sub-stories, one could decide which among the lot of sub-stories require immediate attention. This is particularly useful in applications such as rumour detection where early detection of rumour is important. Categorizing the conversational tweets also into the cluster of the source tweet serves another purpose in this scenario. They help in debugging the truthfulness of a rumour mentioned in the source tweet. For instance, the presence of words such as `incorrect' and `unbelievable' in the reply tweets often indicate that the topic mentioned in the source tweet is not true.

We provide a better measure to evaluate clustering quality in sub-story detection by using adjusted mutual information. We observed that standard story detection approaches such as LSH when applied to sub-story detection task, produced a large number of small sized highly accurate clusters. Standard metrics based on precision favor such clustering approaches but they may not be useful in practice. Though F-score consider recall as well, very high precision often leads to a good F-score. AMI takes into account number of clusters produced by the approach and penalizes those which produces large number of clusters. By correcting agreement between clusters due to chance, AMI measure better reflects the clustering quality of the approaches. We proposed to use it for comparing the quality of clusters produced in the sub-story detection task. HDP performed far better than other approaches in terms of AMI score which makes it an ideal candidate for sub-story detection.

\section{Conclusion}
\label{sect:conclusion}

This paper introduced the sub-story detection task, which  differs from the previously studied  story detection task. 
Secondly, we proposed a probabilistic topic model (hierarchical Dirichlet processes (HDP)) for automatic sub-story detection. HDP performs hierarchical modeling of topics and is effective in modeling sub-stories by learning sub-topics associated with the common topic of the shared real-world event. 

HDP performance was compared against spectral clustering and locality sensitive hashing on several sub-story detection and story detection data sets. In general, we found that SC provides good recall, while LSH provides good precision. HDP, on the other hand, was
found to have balanced precision and recall and achieves the best F-scores on all data sets. This demonstrates that HDP can handle effectively the subtle differences in sub-stories, which leads to an improved clustering performance. The superior performance of HDP is substantiated further by evaluating cluster quality via  adjusted mutual information. 

Lastly, our experiments also demonstrated that considering the conversational structure of tweet threads significantly improved performance of the
sub-story detection approaches.

Future work will address the task of automatic sub-story ranking, which will enable users, such as journalists or emergency responders, to identify and focus on the most important sub-stories within a large volume of tweets surrounding major world events.  

\section*{Acknowledgements}
\label{sect:ack}
We would like to thank Dr. Trevor Cohn for his helpful suggestions. The work is partially supported by the European Union, under grant agreement No. 611233 PHEME.
\clearpage

\bibliographystyle{apalike}
\bibliography{substorydetection_journal}

\end{document}